\documentclass[prd,showpacs,superscriptaddress,twocolumn,
floatfix,preprintnumbers,altaffilletter]{revtex4}

\usepackage{graphicx,psfrag}
\usepackage{mathrsfs}
\usepackage{amsmath,amsfonts,amssymb}
\usepackage{multirow}
\usepackage{comment}
\usepackage{units}
\usepackage{enumerate}
\usepackage[normalem]{ulem} 
\usepackage{longtable}

\usepackage{color}
\definecolor{cyan}{rgb}{0,0.9,0.9}
\definecolor{orange}{rgb}{0.9,0.5,0}
\definecolor{magenta}{rgb}{1,0,1}
\definecolor{purple}{rgb}{0.8,0.4,0.8}
\definecolor{gray}{rgb}{0.8242,0.8242,0.8242}

\begin{document}

\title{Can a black hole-neutron star merger explain GW170817, AT2017gfo, GRB170817A?}

\author{Michael W. Coughlin}
\affiliation{Division of Physics, Math, and Astronomy, California Institute of Technology, Pasadena, CA 91125, USA}

\author{Tim Dietrich}
\affiliation{Nikhef, Science Park 105, 1098 XG Amsterdam, The Netherlands}

\begin{abstract}
The discovery of the compact binary coalescence in both gravitational waves and electromagnetic radiation marks a breakthrough in the field of multi-messenger astronomy and has improved our knowledge in a number of research areas. 
However, an open question is the exact origin of the observables and if one can confirm reliably that GW170817 and its electromagnetic counterparts resulted from a binary neutron star merger. 
To answer the question if the observation of GW170817, GRB170817A, and AT2017gfo could be explained by the merger of a neutron star with a black hole, we perform a joint multi-messenger analysis of the gravitational waves, the short gamma-ray burst, and the kilonova.
Assuming a black-hole neutron star system, we derive multi-messenger constraints for the tidal deformability of the 
NS of $\Lambda > 425$ and for the mass ratio of $q < 2.03$ at 90\% confidence, with peaks in the likelihood near $\Lambda = 830$ and $q = 1.0$.
Overall, we find that a black hole-neutron star merger could explain the observed signatures, however, 
our analysis shows that a binary neutron star origin of GW170817 seems more plausible. 
\end{abstract}


\maketitle


\section{Introduction}

The increasing number of compact binary coalescence detections~\citep{LIGOScientific:2018mvr} by LIGO~\citep{aligo} and Virgo~\citep{avirgo} in their first and second observing runs also increases the hope of detecting black hole-neutron star (BHNS) systems~\citep{Abbott:2016BHNS} in the near future.
BHNS systems have the potential for a joint multi-messenger detection 
of electromagnetic (EM) and gravitational wave (GW) signals~\citep{Bhattacharya:2018BHNS}.
This joint observation would have implications for a number of fields, e.g., 
cosmology, due to reduced distance uncertainties relative to binary 
neutron star detections (BNS) \citep{Nissanke:2009kt,Vitale:2018BHNS}, or nuclear physics, due to constraints on the equation of state (EOS) of matter at supranuclear densities \citep{Pannarale:2011pk}. 

To date, the only multi-messenger observation combining GW and EM signatures was GW170817~\cite{TheLIGOScientific:2017qsa}.
Its electromagnetic counterparts consisted of a short-duration 
gamma ray burst (sGRB), GRB170817A~\cite{Monitor:2017mdv}, 
and its non-thermal afterglow, and a thermal emission (``kilonova'')
at optical, near-infrared, and ultraviolet wavelengths, 
AT2017gfo~\cite{GBM:2017lvd,Arcavi:2017xiz,Coulter:2017wya,Lipunov:2017dwd,
Soares-Santos:2017lru,Tanvir:2017pws,Valenti:2017ngx}. 

While the exact nature of the progenitor system for GW170817 is not 
fully determined, the discovery of a kilonova indicates that the merger 
involved at least one NS. Constraints on the nature of the compact objects from GWs 
can only be drawn under the assumption that the individual spins have been small 
(dimensionless spin below $0.05$) \cite{Abbott:2018wiz}, for which then tidal effects 
suggest that at least one of the compact objects had finite size. 
In addition, GW measurements lead to the conclusion that the second compact object had 
to be of comparable mass~\cite{Abbott:2018wiz}.
Thus, it is possible that this object, while most likely a NS, could have been a 
``light'' black hole (BH)~\cite{Hinderer:2018pei} formed from a prior BNS merger or 
from primordial fluctuations in the early Universe~\cite{GarciaBellido:1996qt}.
Even more exotic, but also possible, is that GW170817 originated from the merger of a neutron star with an exotic compact object, 
e.g., Refs.~\cite{Barack:2018yly,Dietrich:2018jov}. \\

In anticipation of future BHNS detections, there has been a number of studies about the EM and GW signatures arising from a BHNS coalescence. 
The modeling of the GW signal relies on advances in the field of Post-Newtonian Theory~\cite{Blanchet:2013haa,Tagoshi:2014xsa}, 
numerical relativity, e.g.,~\cite{Duez:2008rb,Kyutoku:2010zd,Etienne:2011ea,Kyutoku:2011vz,Foucart:2012vn,Foucart:2013psa,Foucart:2014nda,Kawaguchi:2015bwa,Foucart:2018lhe}, the effective-one-body formalism, e.g.,~\cite{Bernuzzi:2014owa,Hotokezaka:2015xka,
Hinderer:2016eia,Nagar:2018zoe}, and phenomenological waveform modelling, e.g.,~\cite{Lackey:2011vz,Lackey:2013axa,Pannarale:2013uoa,Pannarale:2015jia,Pannarale:2015jka}. 
Modelling of the kilonova signature relies on full-radiative transfer simulations, e.g.,~\cite{Barnes:2016umi,Fernandez:2016sbf,Kasen:2017sxr,Kawaguchi:2018ptg}, or simplified semi-analytical descriptions of the 
observational signatures, e.g.,~\cite{Metzger:2016pju,Kawaguchi:2016ana,Perego:2017wtu,Huang:2018vdq}.

Of central importance for the GW and EM signatures is the final fate of the NS during the merger process. Depending on the mass ratio, the BH's spin, and the EOS, the NS is either torn apart by the tidal forces or plunges directly into the BH~\cite{Shibata:2011jka}.
In the case of a tidally disrupted NS, material is either directly accreted onto the BH, matter forming a disk surrounding the BH~\cite{Foucart:2012nc,Foucart:2018rjc}, 
and material ejected from the system~\cite{Kawaguchi:2016ana}. 
It is this unbound material that yields the processes that power the kilonova~\cite{Metzger:2016pju,Kawaguchi:2016ana}.

In this work, we will study the GW and EM signatures related to GW170817 to understand the origin of the binary. 
Most of the previous analyses assumed a BNS progenitor, e.g., Refs.~\cite{De:2018uhw,Abbott:2018exr,
Radice:2017lry,Radice:2018ozg,Coughlin:2018miv,
Bauswein:2017vtn,Annala:2017llu,Most:2018hfd,Ruiz:2017due,
Margalit:2017dij,Rezzolla:2017aly,Shibata:2017xdx}.
Recently, Hinderer et al.~\cite{Hinderer:2018pei} performed a first joint 
GW and EM analysis of GW170817 as applied to BNS and BHNS mergers with similar masses, 
using bolometric lightcurves to perform the comparison. 
They succeed in ruling out a BHNS merger with mass ratios near to 1
and find generally that only $40\%$ of the GW posterior is compatible with the 
kilonova observation. 
In this paper, we will perform a similar analysis as Ref.~\cite{Hinderer:2018pei}, 
but combining information from three separate sources: GW170817, GRB170817A, and AT2017gfo 
to perform a multi-messenger Bayesian parameter analysis of a potential BHNS merger 
(see \cite{Coughlin:2018fis} for a BNS analysis).
We will derive joint constraints on the binary mass ratio $q$ 
and the tidal deformability $\Lambda$ of the NS.
Finally, we show that our multi-messenger constraints 
lead to a higher chance that GW170817 was produced 
by a BNS and not a BHNS merger in line with the findings of~\cite{Hinderer:2018pei}.

\section{Analyzing GW170817 as a BHNS merger}

\begin{figure}[t]
    \centering
    \includegraphics[width=3.5in]{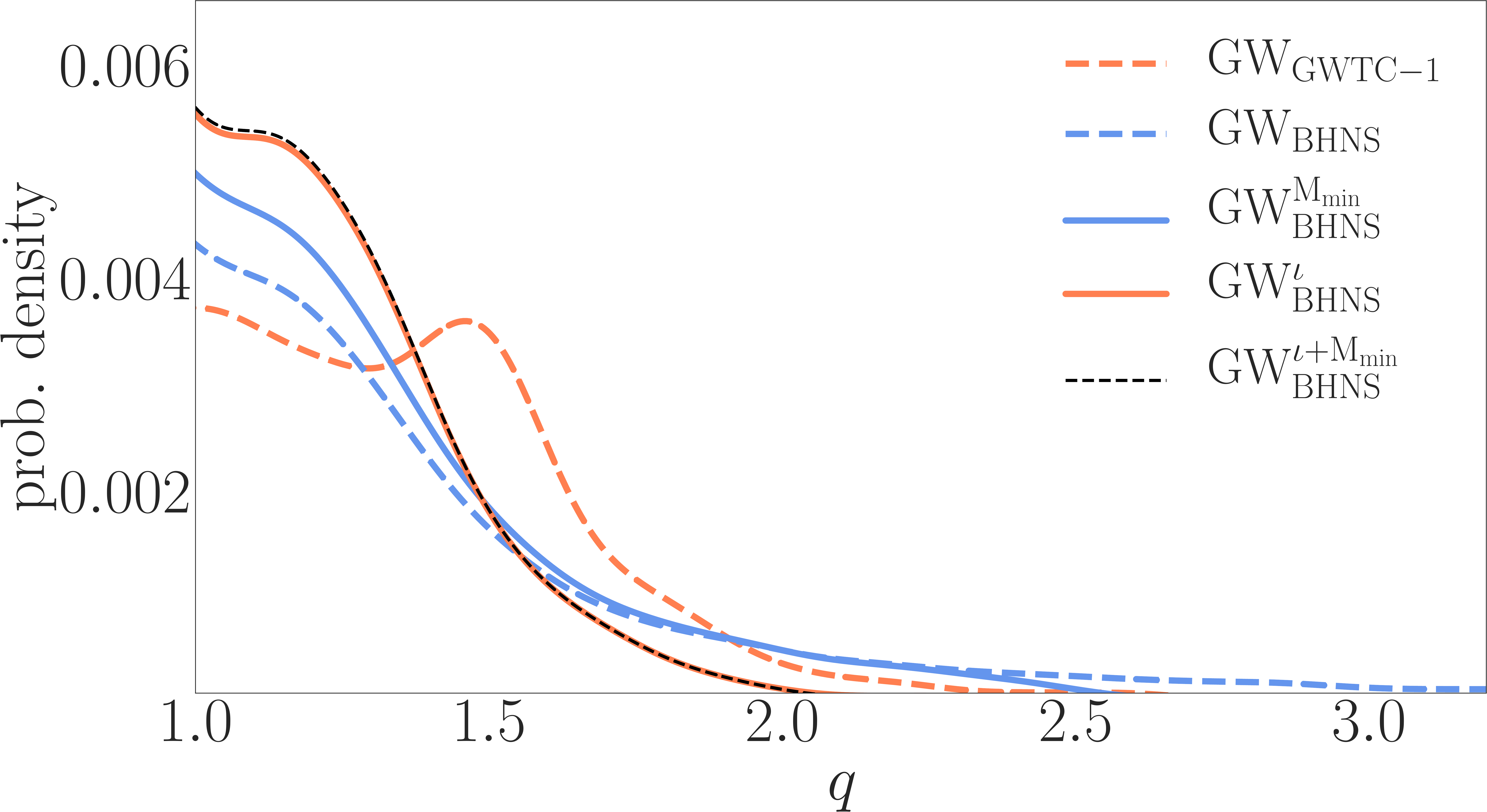}
    \includegraphics[width=3.5in]{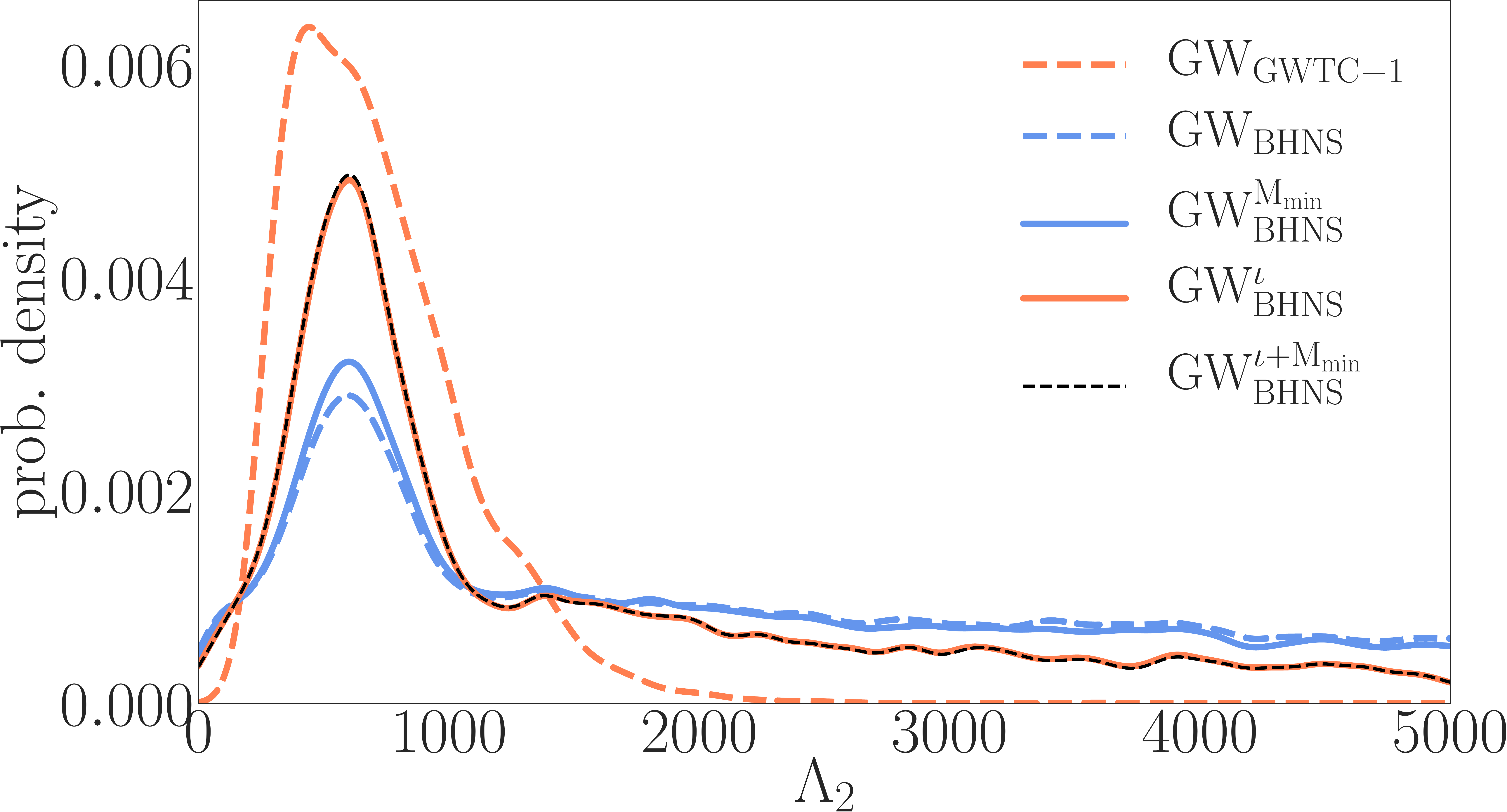}
    \caption{Probability density function obtained from the GW analysis
    for the mass ratio (top panel) and the tidal deformability of the 
    secondary object (bottom panel).
    In addition to showing the GWTC-1 posterior, denoted as '$\rm GW_{GWTC-1}$',
    we show the posterior for an analysis of GW170817 under the 
    assumption that $\Lambda_1 =0$, denoted as '$\rm GW_{\rm BHNS}$'. 
    In addition, we restrict '$\rm GW_{\rm BHNS}$' further to 
    incorporate the viewing angle constraint obtained from GRB170817A ('$\rm GW_{BHNS}^{\iota}$')
    and a minimum NS mass of $M_{\rm}=0.89M_\odot$ ('$\rm GW_{BHNS}^{M_{min}}$'). 
    The final posterior is obtained by a combination of all information, 
    ('$\rm GW_{BHNS}^{\iota+ M_{min}}$').}
    \label{fig:GW}
\end{figure}

Although the GW signal allows estimates of the masses ($m_{1,2}$), 
spins ($\chi_{1,2}=|{\bf S_{1,2}}|/(m_{1,2}^2)$ [in geometric units]), 
and tidal deformabilities ($\Lambda_{1,2}$) of the compact objects, 
the individual quantities are assigned with large uncertainties. 
This is caused by the fact that the GW phase evolution is 
determined mostly by a small number of special combinations of the individual parameter. 
Among these parameters, are the chirp mass,
\begin{equation}
 \mathcal{M} =  M \left( \frac{(1+q)^{2}}{q}\right)^{-3/5},
 \label{eq:Mchirp}
\end{equation}
the effective spin parameter,
\begin{equation}
\chi_{\rm eff_{PN}} = \frac{m_1}{M} {\chi_z}_1 + \frac{m_2}{M} {\chi_z}_2 -
\frac{38}{113} \frac{m_1 m_2}{M^2}({\chi_z}_1 + {\chi_z}_2),
\label{eq:chieff}
\end{equation}
and the tidal deformability,
\begin{equation}
\tilde{\Lambda} = \frac{16}{13} \frac{\Lambda_2+\Lambda_1 q^5 + 12 \Lambda_1 q^4 + 12 \Lambda_2 q}{(1+q)^5},
\end{equation}
which are the main measures with respect to masses, spins, and tides. 
We note that in the case of a BHNS origin of GW170817, 
$\tilde{\Lambda}$ depends only on $q$ and $\Lambda_2$ 
(assuming that the more massive component is a BH).\\

We summarize our main findings with respect to the GW analysis 
in Fig.~\ref{fig:GW} in which we show the mass ratio (top) and 
the tidal deformability of the secondary compact 
object (bottom panel).
We start by presenting results of the first GW transient catalog~\cite{LIGOScientific:2018mvr} 
in which no assumption on the type of the compact binary has been made, i.e., 
the analysis is generic and allows to interpret the system as a BNS, BHNS, 
or even a binary black hole merger. For our purpose, we make use of the high-spin prior results available at 
\texttt{https://dcc.ligo.org/LIGO-P1800370} since there is the chance that
a high spinning BH was present in the system prior to merger; cf.~orange dashed line in Fig.~\ref{fig:GW}.

To focus on a potential BHNS origin, 
we perform a Bayesian analysis of the system using 
the bilby infrastructure~\cite{Ashton:2018jfp} 
employing the IMRPhenomD\_NRTidal~\cite{Dietrich:2018uni} approximant. 
To ensure that we are describing a BHNS system, we set the tidal deformability of 
the primary binary component to zero, i.e, $\Lambda_1=0$,~\footnote{We note that we explicitly 
assume that the more massive component of the binary is a BH and that we do not consider the scenario 
where a BH is bound in a binary with a more massive NS.}. 
To our knowledge, this is the first  Bayesian analysis 
of GW170817 which assumes a BHNS origin of the system. 
Note that although IMRPhenomD\_NRTidal was originally developed for BNS systems, 
a recent comparison with state-of-the-art numerical relativity waveforms 
indicates that for the observed frequencies, the waveform approximant is 
also capable of describing BHNS systems~\cite{Foucart:2018lhe}. 
In our analysis, we make use of the following prior choices, 
$\mathcal{M}  \in (0.87, 1.74) M_\odot$, 
$q  \in (0.125, 1.0) M_\odot$, $\Lambda_2 \leq 7500$, ${\chi_z}_1 \leq 0.95$, and ${\chi_z}_2 \leq 0.05$.
We sample in flat priors over $\mathcal{M}$, $q$, $\Lambda_2$, ${\chi_z}_1$ and ${\chi_z}_2$.
We also use a distance prior of  $40.4\pm 3.4$ \cite{Hjorth:2017a} and find that once $\Lambda_1=0$ is incorporated, 
the constraint on the mass ratio and also on the tidal deformability
of the secondary object are less tight than before (although this could also be related to differences in the waveform models used in these analyses); cf.~dashed blue line in Fig.~\ref{fig:GW}.

We further want to restrict our analysis to 
incorporate additional knowledge from the multimessenger observation, 
namely the detection of GRB170817A and its afterglow, and therefore remove
all samples with viewing angles inconsistent with the GRB analysis 
of \cite{vanEerten:2018vgj}. In more detail, Ref.~\cite{vanEerten:2018vgj} 
finds a viewing angle of  $22\pm 6$, where the error refers to the 1-$\sigma$ uncertainty. 
We increase this to a $2\sigma$ error, i.e., to $22\pm 12$ degree, 
to obtain a more conservative bound on our final results; 
see solid orange line. 

Furthermore, we also include information about the formation scenario 
of NSs and request a minimum NS mass of about $0.89M_\odot.$
The particular value of $0.89M_\odot$ is chosen as the minimum 
mass in the sample of Ref.~\cite{Strobel:1999vn} and seems to be a 
conservative lower bound based on more recent computations, e.g.~\cite{Suwa:2018uni}.
We find that restricting the minimum mass of the NS leads to a less significant constraint 
than the restriction of the inclination angle. 

Employing both, the inclination angle and the minimum mass constraint, 
leads to the dashed black line and our final GW result. 

\begin{figure}[t]
  \includegraphics[width=3.5in]{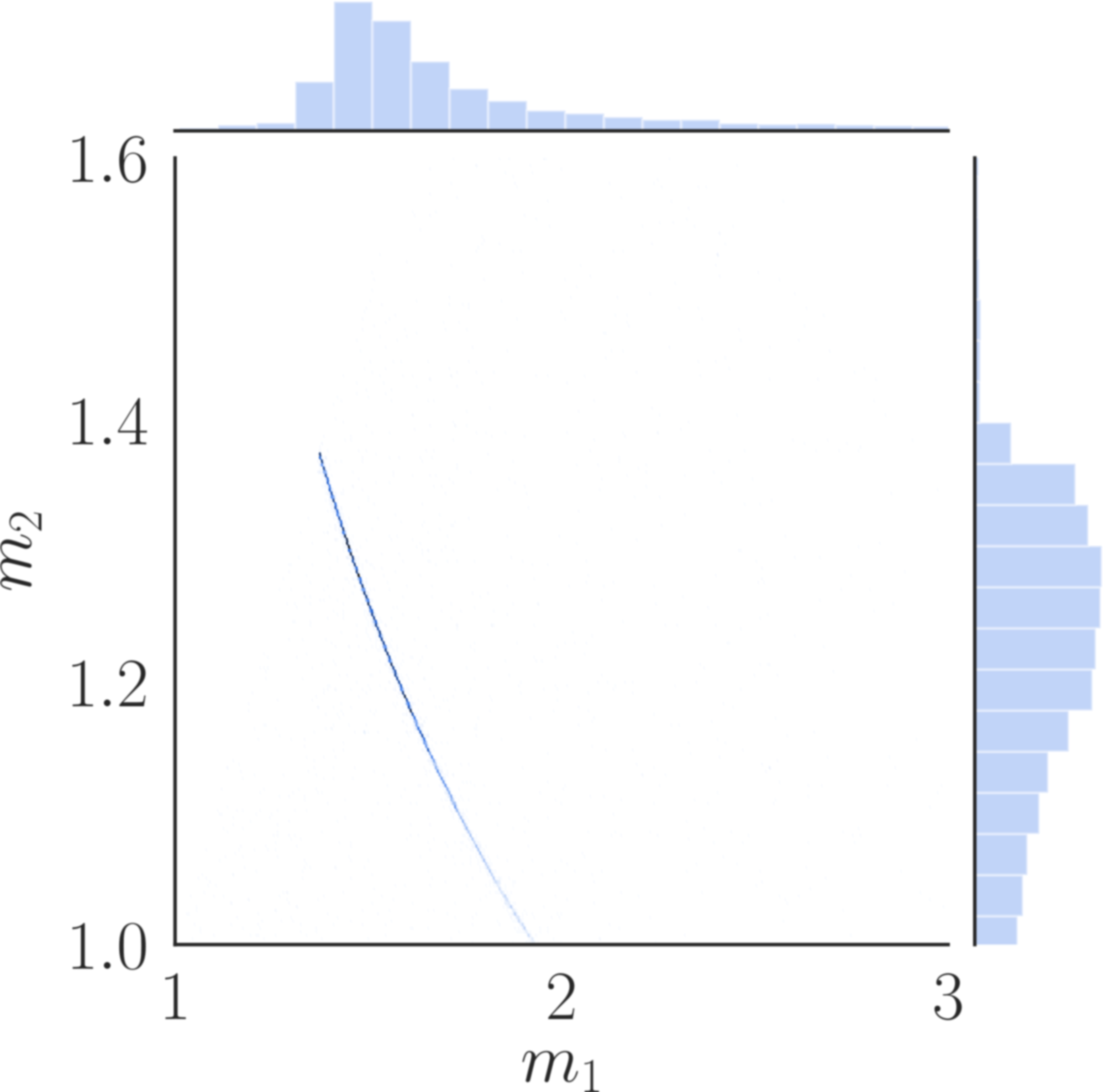} 
  \caption{$m_1$ and $m_2$ under the assumption of a BHNS merger.} 
\label{fig:m1m2}
\end{figure}

Overall, this procedure results in constraints (90\% upper bounds) 
of $q \leq 1.59$ and $\Lambda_2 \leq 3564$.
We also show for this final result the posteriors for $m_1$ and $m_2$ 
in Figure~\ref{fig:m1m2}.

\section{AT2017gfo and GRB170817A arising from a BHNS merger?}

We now jointly analyze the EM data from AT2017gfo and GRB170817A under 
the assumption that they were produced from a BHNS merger.\\

\textbf{The kilonova AT2017gfo:}
We first fit the observational data of AT2017gfo~\cite{Coughlin:2018miv,Smartt:2017fuw,GBM:2017lvd} 
with the 2-component radiative transfer model of Kasen et al.~\cite{Kasen:2017sxr}. 
These kilonova models are parameterized by the ejecta mass 
$m_{\rm ej}$, the lanthanide mass fraction $X_{\rm lan}$ 
(related to the initial electron fraction), and the ejecta velocity $v_{\rm ej}$ 
of each component.
We combined these models with a Gaussian Process Regression 
framework~\cite{Coughlin:2018miv} to obtain information about the ejecta from the lightcurves.
We note that the analysis is subject to possible systematic errors 
arising from approximations such as the spherical geometry of the ejecta and 
the non-inclusion of mixing between the two ejecta components \cite{Fernandez:2016sbf}.

The use of two components is motivated by the different processes 
contributing to the kilonova. Broadly, these are known as {dynamical ejecta}, 
generated in the merger process by tidal torques, and {disk winds ejecta}, 
which result from neutrino energy, magnetic fields, viscous evolution and/or 
nuclear recombination (e.g.~\cite{Kohri:2005tq,Surman:2005kf,Metzger:2008av,
Siegel:2014ita,Just:2014fka,Rezzolla:2014nva,Song:2017qyc}). 
We associate the first component with 
dynamical ejecta and the second with the disk wind.

For the dynamical ejecta, we use the fits of Kawaguchi et al.~\cite{Kawaguchi:2016ana} 
to tie the binary parameters to those of the ejecta. 

Kawaguchi et al.~\cite{Kawaguchi:2016ana} shows that the ejecta mass and 
velocity can be approximated by: 
\begin{small}
\begin{eqnarray}
	\frac{M_{\rm ej}}{M_{\rm NS,*}}& = 
	&{\rm Max}\left\{a_1 q^{n_1}\frac{1-2 C}{C}-a_2\,q^{n_2}\,{\hat R}_{\rm ISCO} (\chi_{\rm BH}) \right. 
	\label{eq:BHNS:mej_fit} \nonumber \\
	 &+& \left. a_3\left(1-\frac{M_{\rm NS}}{M^*_{\rm NS}}\right)+a_4,0 \right\},   \\
	v_{\rm ej}&=& b_1 \,q+b_2 \label{eq:BHNS_vfit}.
\end{eqnarray}
\end{small}
where $a_1,a_2,a_3,a_4,n_1,n_2,b_1,b_2$ are fitting parameters, see Ref.~\cite{Kawaguchi:2016ana} for details.
It uses the normalized ISCO radius $\hat{R}_{\rm ISCO} = R_{\rm ISCO}/M_{\rm BH}$, where
for $q\rightarrow \infty$, the ISCO radius becomes 
\begin{equation}
\hat{R}_{\rm ISCO} = 3+Z_2-{\rm sgn}(\chi_{\rm BH})\sqrt{(3-Z_1)(3+Z_1+2Z_2)},\qquad
\end{equation}
where
\begin{equation}
Z_1=1+(1-\chi_{\rm BH}^2)^{1/3}[(1+\chi_{\rm BH})^{1/3}+(1-\chi_{\rm BH})^{1/3}],
\end{equation}
and
\begin{equation}
Z_2=\sqrt{3\chi_{\rm BH}^2+Z_1^2}. 
\end{equation}
The chirp mass, Eq.~\eqref{eq:Mchirp}, $\mathcal{M}=1.186M_{\odot}$ measured by the GW 
inference allows us to relate the mass ratio directly to the mass of the NS, $M_{\rm NS}$.
Furthermore, also for our EM analysis, we make use of the 
minimum NS mass of $0.89M_\odot$. 

We note that the phenomenological
relation of~\cite{Kawaguchi:2016ana} is calibrated to numerical relativity simulations with mass 
ratio $q\geq3$ and will be used in this work outside its calibration region. 
However, by construction, the relation is bound and shows no artificial 
behavior for smaller mass ratios. 
Nevertheless, we point out that due to the usage of Eqs.~\eqref{eq:BHNS:mej_fit} 
and~\eqref{eq:BHNS_vfit} outside its calibration region we might be effected 
by systematic uncertainties, which are unable to quantify at the current stage. 
To improve the phenomenological relations a much larger set of numerical relativity 
simulations is needed, which currently is not 
available within the numerical relativity community. 

To connect the gravitational and baryonic mass to the 
compactness, we employ the quasi-universal relation presented in Ref.~\cite{Coughlin:2017ydf}:
\begin{equation}
 \frac{M^*}{M}= 1 + a~C^n, \label{eq:quasi_univ_fit}
\end{equation}
with $a=0.8858$ and $n=1.2082$. 

Following Coughlin et al.~\cite{Coughlin:2018fis}, we assume that the dynamical ejecta is proportional to the total first component:  
\begin{equation}
    m_{\rm ej, 1} = \frac{1}{\alpha} \  m_{\rm dyn}, \qquad v_{\rm ej, 1} = v_{\rm dyn}.
\label{eq:dyn}
\end{equation}
where we sample over a flat prior in $\alpha$, which encodes this fraction. 
We sample directly in $\Lambda_2$, and compute the compactness of the NS by 
$C = 0.371 - 0.0391 \log(\Lambda_2) + 0.001056 \log(\Lambda_2)^2$.

We now turn to the second ejecta component.
The baryon mass remaining outside the resulting BH after merger, 
known as the debris disk mass $m^{\rm disk}$, 
determines the mass available for the counterparts. 
Ref.~\cite{Foucart:2018rjc} provides a prediction of the disk mass as a function of 
the NS's compactness $C$, the dimensionless BH spin $\chi_{\rm BH}$, 
and the mass ratio $q$:
\begin{equation}
\hat{M}^{\rm disk}_{\rm model} = \left[{\rm Max}\left(\alpha \frac{1-2C}{\eta^{1/3}} - 
\beta \hat{R}_{\rm ISCO}\frac{C}{\eta}+\gamma,0\right)\right]^\delta \ ,
\label{eq:foucart}
\end{equation}
with $\alpha=0.406, \beta=0.139, \gamma=0.255$, $\delta=1.761$ and
\begin{equation}
\eta = \frac{m_1 m_2}{(m_1+m_2)^2}
\end{equation}
being the symmetric mass ratio.
Eq.~\eqref{eq:foucart} uses numerical relativity data covering 
regions of the parameter space including 
comparable masses and high BH spins, 
$q\in [1,7]$, $\chi_{\rm BH}\in[-0.5, 0.97]$, 
and $C\in [0.13,0.182]$, cf.~\cite{Etienne:2008re,Foucart:2010eq,
Foucart:2011mz,Kyutoku:2011vz,Foucart:2019bxj}. 
The covered parameter space in $C$ corresponds to $\Lambda_2 \in [304,2469]$ 
if we employ the quasi-universal relation mentioned above. 
We point out that as for the dynamical ejecta description, part of our analysis covers 
regions outside the calibration region of Eq.~\eqref{eq:foucart}. Furthermore, 
based on the limited number of numerical relativity simulations, 
systematic biases of the relations connecting estimated ejecta and disk masses with 
binary properties are relatively uncertain. 
Part of this uncertainty is incorporated due to the large modelling uncertainty of 1 magnitude 
errors, but more simulations and additional work is needed for a better quantitative assessment of the modelling 
uncertainties. 

Similar to the analysis for the first component, we assume for the second ejecta component
\begin{equation}
    m_{\rm ej, 2} = \zeta \ m_{\rm disk}, 
\label{eq:wind}
\end{equation}
i.e., only a fraction of the disk is ejected. 
We restrict $\zeta$ to lie within $\zeta \in [0,0.5]$ 
as for the analysis in~\cite{Coughlin:2018fis}.

Fitting the observational data of 
AT2017gfo~\cite{Coughlin:2018miv,Smartt:2017fuw,GBM:2017lvd}
yields posteriors for $m_{\rm ej, 1}, v_{\rm ej, 1},$ $X_{\rm lan, 1}$, 
$m_{\rm ej, 2}, v_{\rm ej, 2},$ and $X_{\rm lan, 2}$
(please see Figs.~5 and 6 of Ref.~\cite{Coughlin:2018fis}).

\begin{figure}
    \centering
    \includegraphics[width=3.5in]{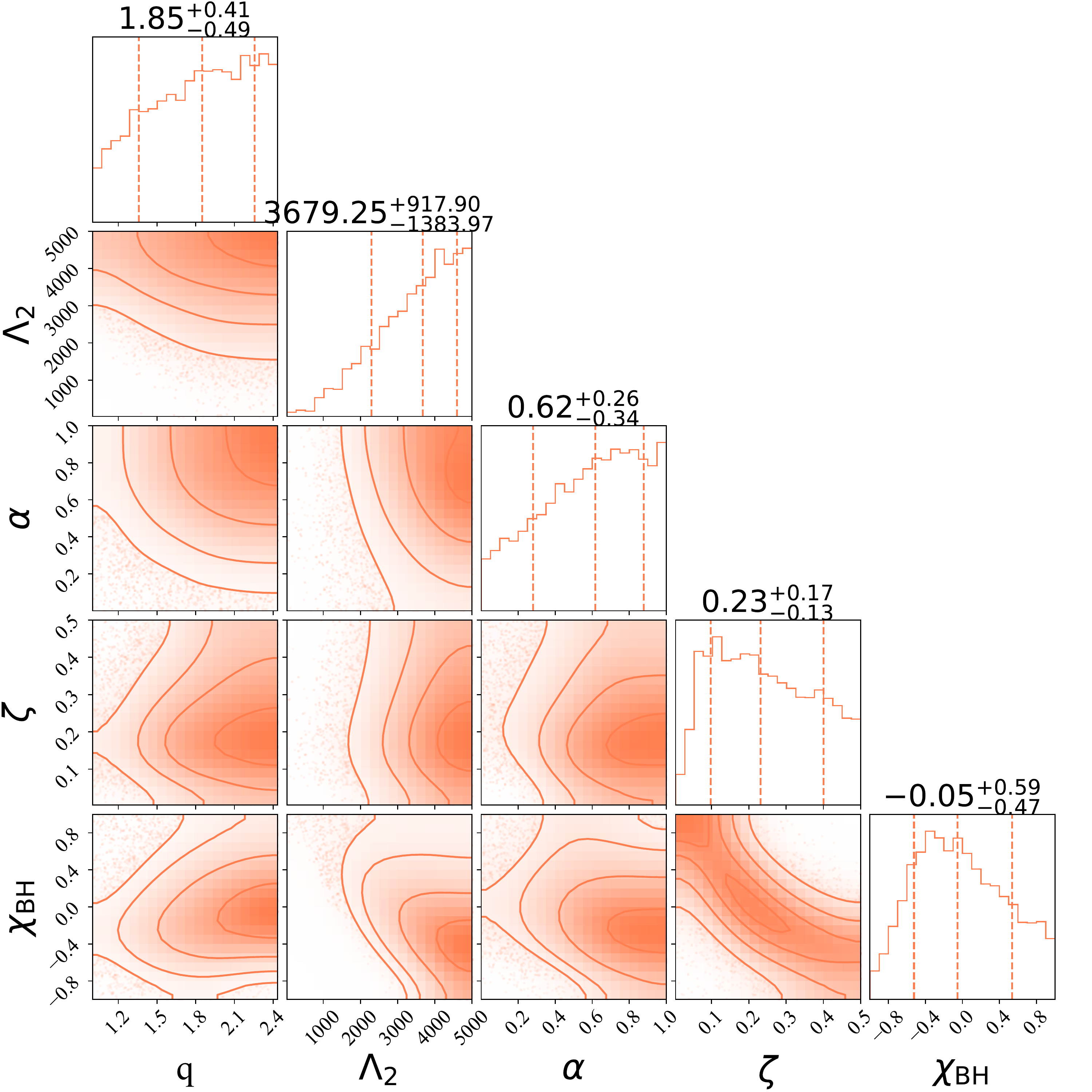} 
    \caption{Posterior distributions for our analysis of AT2017gfo. 
    We present posteriors for the mass ratio $q$, the tidal deformability of the NS
    $\Lambda_2$, the fraction of the first ejecta component 
    related to dynamical ejecta $\alpha$, the fraction of the disk mass ejected 
    as the second component ejecta $\zeta$, and 
    the BH spin parameter $\chi_{\rm BH}$.}
    \label{fig:EM1}
\end{figure}

Based on Eqs.~\eqref{eq:dyn} and~\eqref{eq:wind}, we use a 
Kernel Density Estimator (KDE) to compare the predictions with the fits 
from the two-component kilonova data, 
yielding constraints on $q, \Lambda_2, \alpha, \zeta, \chi_{\rm BH}$.
These can be identified in Fig.~\ref{fig:EM1} (please see Fig. 6 of Ref.~\cite{Coughlin:2018fis} for similar plot in the BNS case).
We find that as equal mass ratio systems are relatively unlikely. 
The 50th percentile lies at $q=1.85$. 
Considering the tidal deformability, we find that smaller 
values of $\Lambda_2 <1000$ are unlikely based on our analysis
and that the posterior seems to rail against the prior boundary of 
$\Lambda_2=5000$, which we impose to be consistent 
with the upper boundary used by the LIGO and Virgo analysis in~\cite{Abbott:2018wiz}. 
In fact, allowing for even larger values of $\Lambda_2$ leads to a posterior 
distribution peaking around $5000$, which we note lies above the calibration 
region of the NR fits. 
As might be expected for BHNS mergers, which generally have larger predictions for dynamical ejecta, $\alpha$ peaks near the top end of the prior at $\alpha=1$, with less support at lower values.
$\chi_{\rm BH}$ has most of its support at positive values, peaking near $\chi_{\rm BH} = 0.25$, which arises from negative values resulting in smaller values of the dynamical ejecta.
Similarly, $\zeta$ peaks at lower values near $\zeta=0.1$ with less support at the top end of 
the prior, indicating a smaller contribution from the disk.\\

\begin{figure}
    \centering
    \includegraphics[width=3.5in]{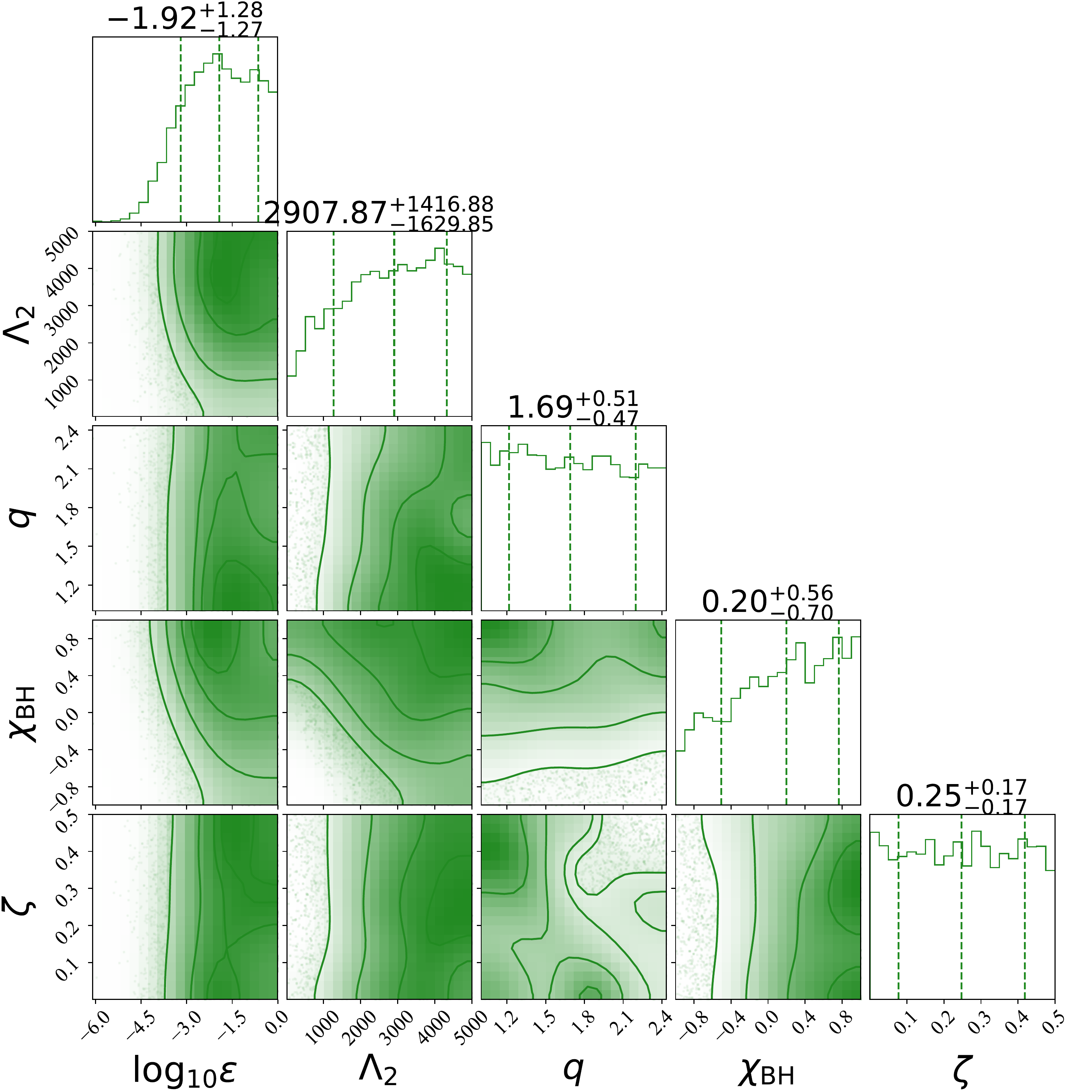}
    \caption{    
    Posterior distributions for the GRB analysis showing 
    constraints on the mass ratio $q$, the tidal deformability of the NS
    $\Lambda_2$, the fraction of the disk mass ejected 
    as the second component ejecta $\zeta$, the effective spin parameter $\chi_{\rm BH}$, 
    and the fraction of the disk rest mass converted to trigger the sGRB $\epsilon$.}
    \label{fig:EM2}
\end{figure}

\textbf{The gamma-ray-burst GRB170817A:}
In the next step, we use the results obtained from the analysis of AT2017gfo 
and combine it with energy constraints obtained from the observation of 
GRB170817A~\cite{vanEerten:2018vgj}.
To do so, we assume that the GRB is powered by the accretion 
of matter from the debris disk onto the BH~\cite{Eichler:1989ve,Paczynski:1991aq,Meszaros:1992ps,Narayan:1992iy}.
Tying this into the kilonova analysis means that the energy is proportional 
to the disk mass minor the part of the baryonic mass which gets ejected by winds, i.e., 
\begin{equation}
 E_{jet} = \varepsilon ( m_{\rm disk} - m_{ej,2} ) = \varepsilon m_{\rm disk} (1 - \zeta).
 \label{eq:EGRB}
\end{equation}
In the BNS analysis~\cite{Coughlin:2018fis}, 
we used three different fits to the GRB afterglow: 
van Eerten et al~\cite{vanEerten:2018vgj}, Wu and MacFadyen~\cite{Wu:2018bxg}, 
and Wang et al~\cite{Wang:2018nye}.
We showed that our analysis was robust against potential systematic uncertainties 
by checking the consistency between the three different GRB analyses. For this reason, 
we will here only adopt the model of van Eerten et al.~\cite{vanEerten:2018vgj}.
Ref.~\cite{vanEerten:2018vgj} used a Gaussian structured form of the jet and constrained the energy in 
the jet to be $\log_{10} [E_{\rm jet}/{\rm erg}] = 50.30^{+0.84}_{-0.57}$. 

We make use of the posteriors of $\zeta, \Lambda_2, q$, and $\chi$ 
from the kilonova analysis as priors for the GRB analysis.
The analysis proceeds by comparing the estimated energy 
from~\cite{vanEerten:2018vgj} to the energy estimated in equation~\eqref{eq:EGRB}.
Final posteriors are shown in Fig.~\ref{fig:EM2} (please see Fig. 7 of Ref.~\cite{Coughlin:2018fis} for similar plot in the BNS case).
As compared to the kilonova posteriors, the analysis more strongly disfavors higher mass ratios, which generally leads to smaller disk masses inconsistent with the second component.
Similarly, higher values for the effective spin are preferred, which leads to larger disk masses, although a negative spin is not ruled out.
The posteriors for $\zeta$ and $\Lambda_2$ are not changed significantly compared to the kilonova-based results.

\section{Discussion}

\begin{figure*}[t]
    \centering
    \includegraphics[width=3.5in]{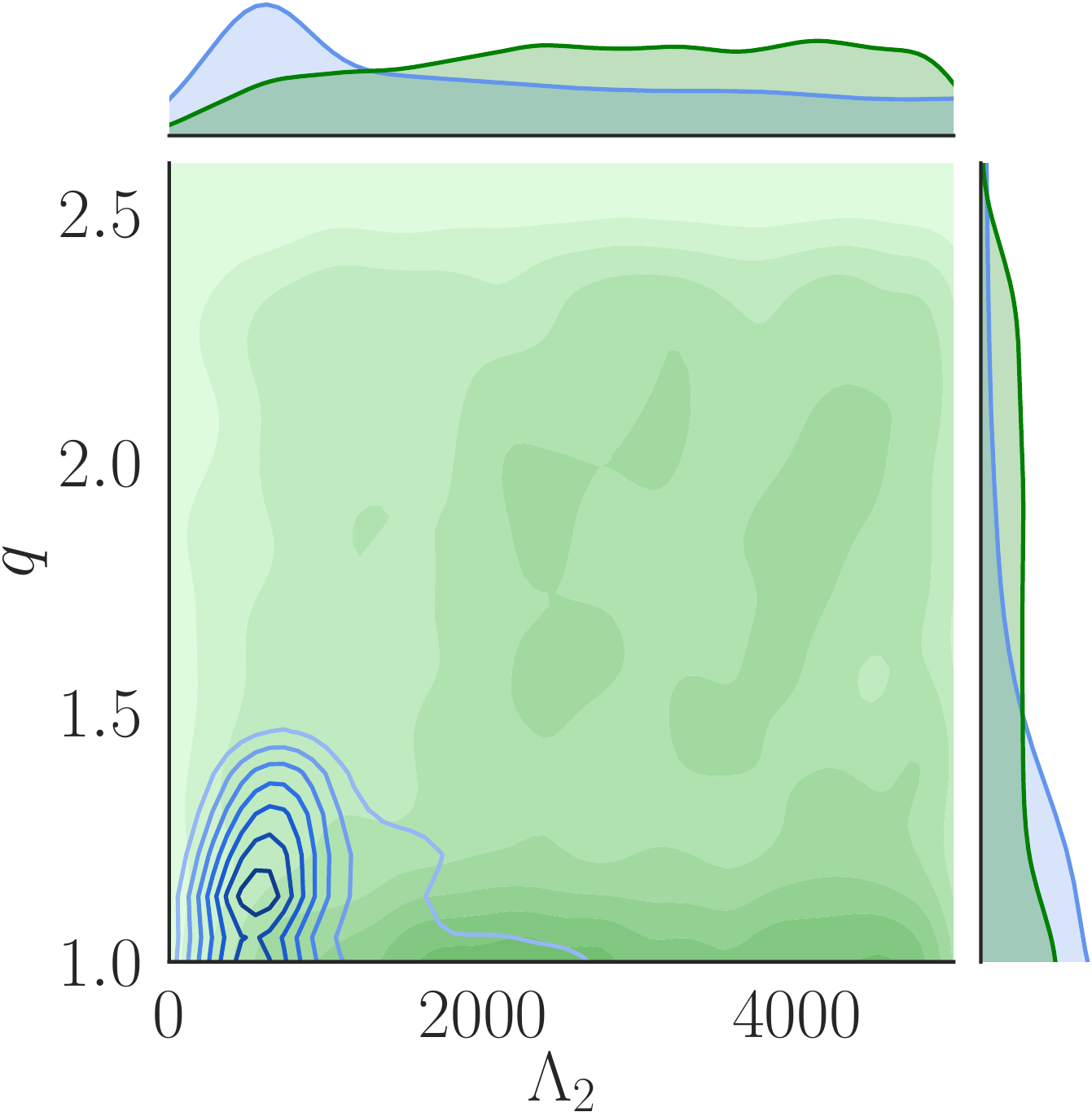}
    \includegraphics[width=3.5in]{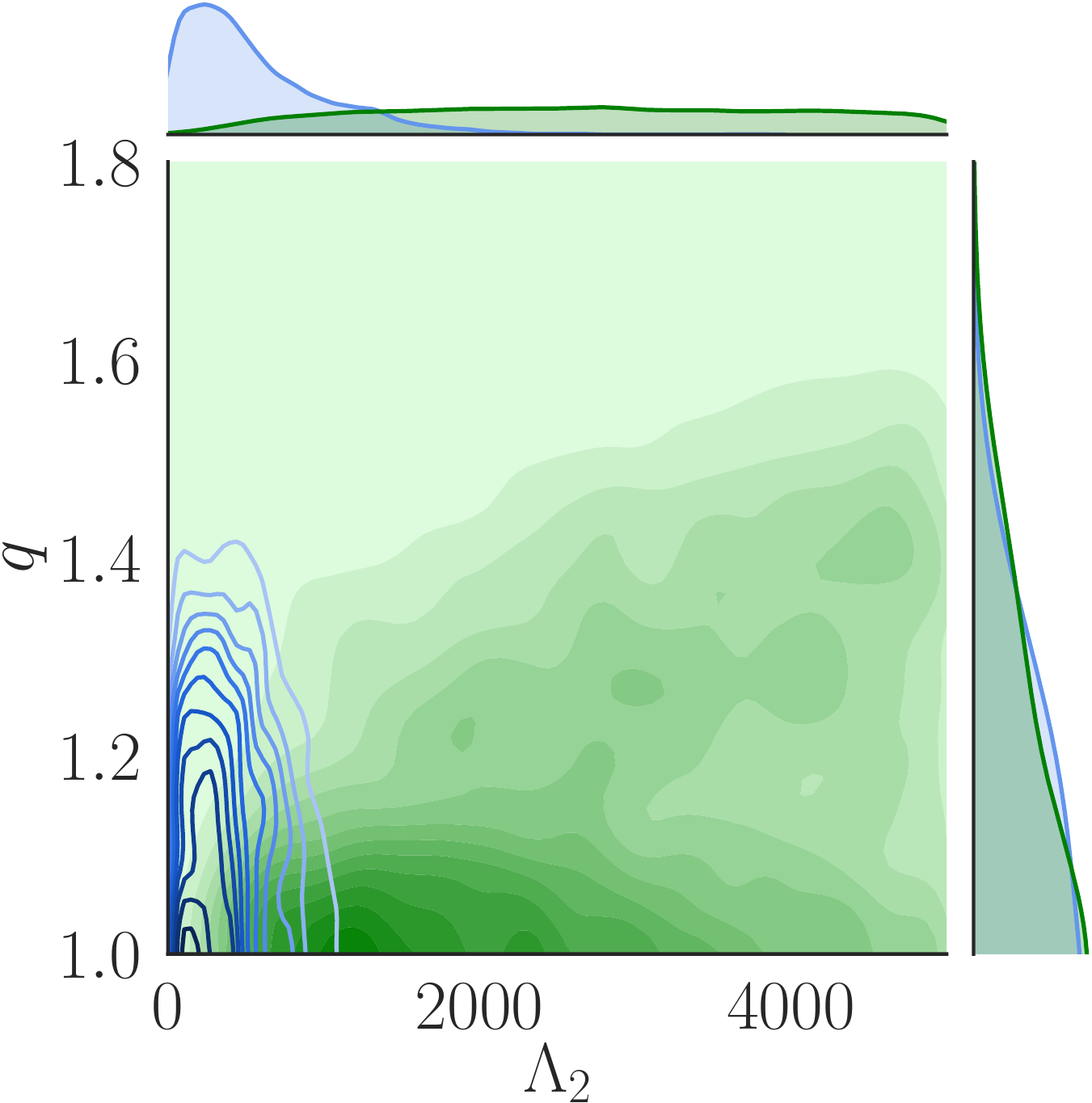}
    \caption{On the left is the $\Lambda_2$-$q$ posterior distribution for the GW (blue) and 
    EM (green) analysis. The EM posterior refers to the results obtained from 
    the analysis of GRB170817A using the results of AT2017gfo as input priors.
    On the right is the same assuming that GW170817/AT2017gfo/GRB170817A 
    arose from the merger of two neutron stars.}
    \label{fig:GW+EM}
\end{figure*}

\begin{figure*}[t]
 \includegraphics[width=3.5in]{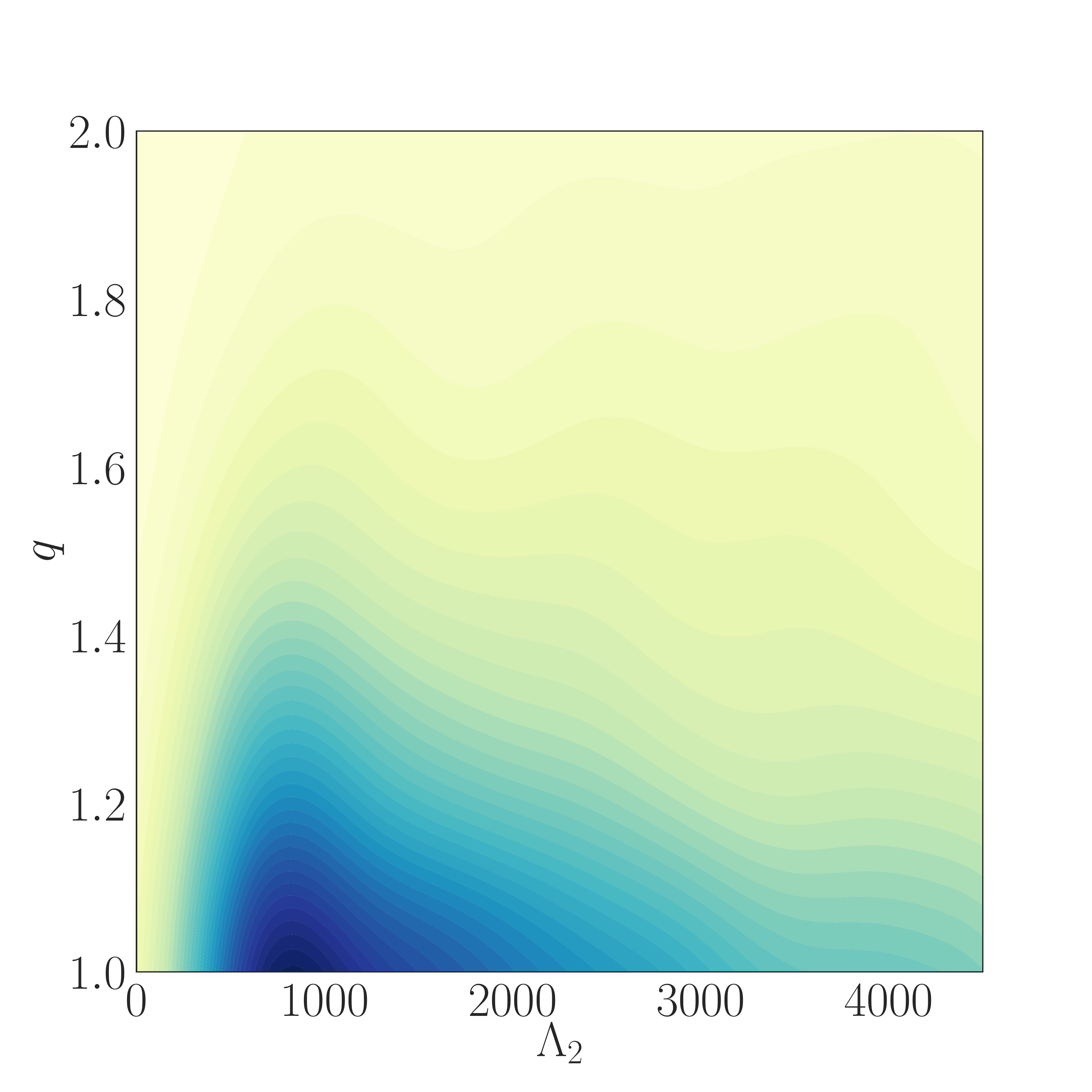} 
 \includegraphics[width=3.5in]{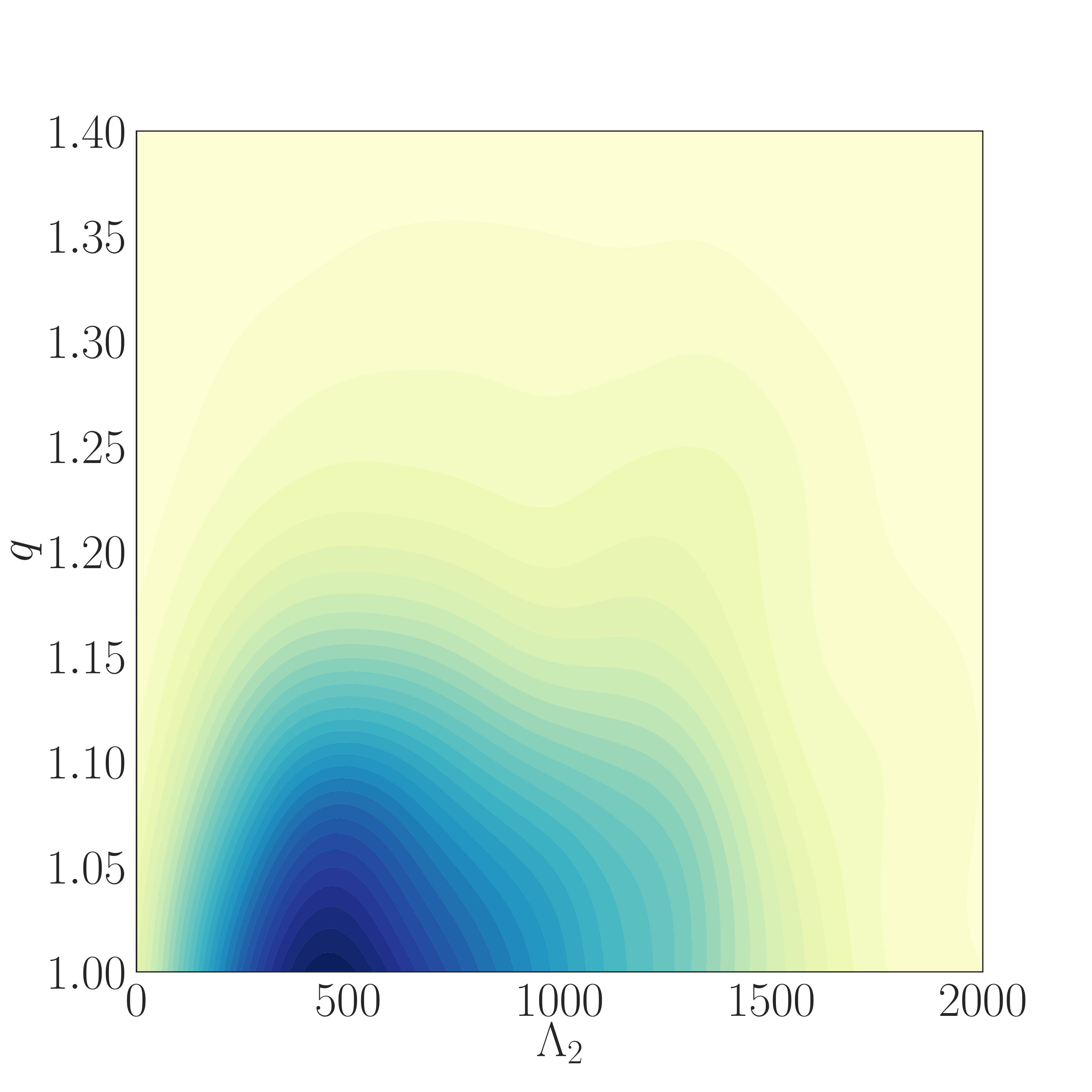} 
  \caption{On the left are the joint constraints on the mass ratio $q$ and tidal deformability of the NS $\Lambda_2$ under the assumption of a BHNS merger. The analysis favors almost equal mass 
  binaries and NS tidal deformabilities of 
  $\Lambda_2\approx 1000$, i.e., a NS radius of $R\approx 14\rm km$.
  On the right are the same under the assumption of a BNS merger.
  Note that both panels employ different axis ranges.} 
\label{fig:joint}
\end{figure*}

\textbf{A combined multi-messenger astronomy constraint:}
We now combine the GW and EM observations of GW170817 to make joint constraints on a potential BHNS binary. 
We directly take the posterior distributions for $q$ and $\Lambda_2$  obtained from the GW analysis in Figure~\ref{fig:GW}.
Binning these results yields the posterior distribution of $q$ and $\Lambda_2$ 
for a BHNS progenitor of GW170817. 

Figure~\ref{fig:GW+EM} shows the $q$-$\Lambda_2$ posterior for the GW (blue) and EM (green) 
analysis. We can now construct a joint distribution for $\Lambda_2$ and $q$ by 
multiplying the probability distributions for
\begin{equation}
 P_{\rm MMA}(\Lambda,q) = P_{\rm GW}(\Lambda_2,q) \times P_{\rm EM}(\Lambda_2,q)  \times Pr(\Lambda_2) Pr(q). \label{eq:PMMA}
\end{equation}
where the contributions from the priors are encoded by $Pr(\Lambda_2)^2 Pr(q)^2$ (and we are implicitly setting the prior on the data to be 1).
We remind the reader that these priors are flat over the bounds considered.
We show the joint constraints on the binary parameters and EOS in Figure~\ref{fig:joint}.
In general, the combined samples of both analyses are consistent with 
almost equal mass systems with large values of $\Lambda_2$. 
More quantitatively, this analysis results in a constraints on 
the tidal deformability of the NS of $\Lambda > 425$ and on the mass 
ratio of $q < 2.03$ at 90\% confidence, 
with peaks in the likelihood near $\Lambda = 830$ and $q = 1.0$.

\textbf{Comparison of BHNS and BNS:}
To contrast with our BHNS analysis, we present a possible scenario 
in which GW170817 and its EM counterparts arose from the merger of two NSs.
In the following, we point out differences between the BHNS and the BNS analysis. 
See also the detailed explanation in Ref.~\cite{Coughlin:2018fis} about the BNS analysis. 

For the GW analysis, instead of restriction the tidal deformability of the 
more massive object to be zero and sampling in directly in $\Lambda_2$, 
we employ the posterior samples provided in~\cite{PhysRevD.99.083016}.
For the kilonova analysis of a BNS system, we assume that the first 
component ejecta (dynamical ejecta) can be described by the 
phenomenological fit presented in~\cite{Coughlin:2018fis} 
(Eq.~(2)) and based on~\cite{Dietrich:2016fpt} and that the 
second ejecta component (disk wind ejecta) are 
described by Eq.~(1) of~\cite{Coughlin:2018fis}
and based on~\cite{Radice:2018pdn}.
Eq.~(1) of~\cite{Coughlin:2018fis} is also employed for the description of the 
debris disk mass used as a central engine for the sGRB. 

Note that the difference between the BNS analysis presented here and 
in Coughlin et al.~\cite{Coughlin:2018fis} is that we do not sample in $\tilde{\Lambda}$, 
but in $\Lambda_2$ with a prior of $\Lambda_2 \in [0,5000]$. 
We do this to allow a direct comparison between the BNS and 
BHNS scenario and to reduce possible systematic biases. 
Furthermore, we assume a maximum NS mass of $M\approx2.17M_\odot$
as proposed in~\cite{Margalit:2017dij}.

\textbf{Probability of a BNS or BHNS merger:}
We want to finish by testing the consistency between the probability 
distributions for the GW and the EM analyses, Fig.~\ref{fig:GW+EM}.
To do so, we assume the parameter estimation analyses are independent from one another 
and compute a Bayesian evidence for each analysis.
To compare the two,  we compute a Bayes factor, which is the ratio between evidences for both analyses.
Formally, 
\begin{equation}
K =  \frac{P_{\rm BNS}}{P_{\rm BHNS}} = \frac{\int_{\Lambda_2} \int_q P_{\rm GW-BNS}(\Lambda_2,q) \times P_{\rm EM-BNS}(\Lambda_2,q)}{\int_{\Lambda_2} \int_q P_{\rm GW-BHNS}(\Lambda_2,q) \times P_{\rm EM-BHNS}(\Lambda_2,q)}.  
\end{equation}
where, because the priors are the same, they divide out in this analysis.
We find that the Bayes factor for the BNS vs. the BHNS case is 3.0. Thus, it seems more likely that the origin of GW170817, AT2017gfo, 
and GRB170817A was a BNS merger~\footnote{We want to mention that a drawback of our method 
is the missing information about the lanthanide or electron fraction in the 
numerical relativity description of the ejecta properties. 
We suggest that an inclusion of those would decrease the probability 
that GW170817 and its EM counterparts are sourced by a BHNS system, 
see e.g.~\cite{Siegel:2017nub,Siegel:2017jug,Fernandez:2018kax}.}. \\

\textbf{Summary:}
We have used a combined analysis of GW170817, GRB170817A, and AT2017gfo to 
constrain the possibility of the GW and EM signals arising from a BHNS merger. 
To connect the EM signature to the binary properties, we have employed phenomenological 
relations connecting the dynamical ejecta mass and the disk mass to 
the properties of the binary system. Under the assumption that, in particular, 
the relation of~\cite{Kawaguchi:2016ana} can be employed outside its calibration region 
and that GW178017 and its EM counterparts are caused by a BHNS merger, 
we find that the system would have a mass ratio of $q < 2.03$ with a 
non-compact NS of $\Lambda > 425$. 
We compared the BHNS scenario with a BNS scenario and find that the EM and 
GW posteriors have a Bayes factor of 3.0, indicating a BNS system is more 
likely compared to a BHNS system; cf.~also~\cite{Hinderer:2018pei}.

As both GW and EM models improve in the coming years, these types of analyses will 
be useful to further classify the origin of observed multi-messenger structures. 
In particular, improvements in the light curve modeling, such as incorporating viewing angle effects, will be required. 
In addition, GW measurements of a post-merger signal or tidal 
disruption will place further constraints on the progenitor properties.

\acknowledgments

We that Tanja Hinderer, Ben Margalit, Brian Metzger, and 
Samaya Nissanke for helpful discussions and comments on the manuscript. 
Furthermore, we thank Zack Carson, Katerina Chatziioannou, Carl-Johan Haster, Kent Yagi, Nicolas Yunes
for providing us their posterior samples analyzing GW170817 under the assumption 
of a common EOS~\cite{Carson:2019rjx}.
MWC is supported by the David and Ellen Lee Postdoctoral Fellowship at the 
California Institute of Technology. 
TD acknowledges support by the European Union's Horizon 2020 research and innovation 
program under grant agreement No 749145, BNSmergers. 
The authors thank the maintainers of the LIGO data grid for computers used to perform the analysis.

\section*{References}
\bibliographystyle{iopart-num}
\bibliography{refs}

\providecommand{\newblock}{}
\begin{thebibliography}{10}
\expandafter\ifx\csname url\endcsname\relax
  \def\url#1{{\tt #1}}\fi
\expandafter\ifx\csname urlprefix\endcsname\relax\def\urlprefix{URL }\fi
\providecommand{\eprint}[2][]{\url{#2}}

\bibitem{LIGOScientific:2018mvr}
Abbott B~P {\em et~al.\/} (LIGO Scientific, Virgo) 2018  (\textit{Preprint}
  \eprint{1811.12907})

\bibitem{aligo}
{J Aasi et al} 2015 {\em Classical and Quantum Gravity\/} {\bf 32} 074001
  \urlprefix\url{http://stacks.iop.org/0264-9381/32/i=7/a=074001}

\bibitem{avirgo}
{F Acernese et al} 2015 {\em Classical and Quantum Gravity\/} {\bf 32} 024001
  \urlprefix\url{http://stacks.iop.org/0264-9381/32/i=2/a=024001}

\bibitem{Abbott:2016BHNS}
{Abbott et al} 2016 {\em The Astrophysical Journal Letters\/} {\bf 832} L21
  \urlprefix\url{http://stacks.iop.org/2041-8205/832/i=2/a=L21}

\bibitem{Bhattacharya:2018BHNS}
{Bhattacharya} M, {Kumar} P and {Smoot} G 2018 {\em arXiv e-prints\/}
  (\textit{Preprint} \eprint{1809.00006})

\bibitem{Nissanke:2009kt}
Nissanke S, Holz D~E, Hughes S~A, Dalal N and Sievers J~L 2010 {\em Astrophys.
  J.\/} {\bf 725} 496--514 (\textit{Preprint} \eprint{0904.1017})

\bibitem{Vitale:2018BHNS}
Vitale S and Chen H~Y 2018 {\em Phys. Rev. Lett.\/} {\bf 121}(2) 021303
  \urlprefix\url{https://link.aps.org/doi/10.1103/PhysRevLett.121.021303}

\bibitem{Pannarale:2011pk}
Pannarale F, Rezzolla L, Ohme F and Read J~S 2011 {\em Phys.Rev.\/} {\bf D84}
  104017 (\textit{Preprint} \eprint{1103.3526})

\bibitem{TheLIGOScientific:2017qsa}
Abbott B~P {\em et~al.\/} (Virgo, LIGO Scientific) 2017 {\em Phys. Rev.
  Lett.\/} {\bf 119} 161101 (\textit{Preprint} \eprint{1710.05832})

\bibitem{Monitor:2017mdv}
Abbott B~P {\em et~al.\/} (Virgo, Fermi-GBM, INTEGRAL, LIGO Scientific) 2017
  {\em Astrophys. J.\/} {\bf 848} L13 (\textit{Preprint} \eprint{1710.05834})

\bibitem{GBM:2017lvd}
Abbott B~P {\em et~al.\/} (GROND, SALT Group, OzGrav, DFN, DES, INTEGRAL,
  Virgo, Insight-Hxmt, MAXI Team, Fermi-LAT, J-GEM, RATIR, IceCube, CAASTRO,
  LWA, ePESSTO, GRAWITA, RIMAS, SKA South Africa/MeerKAT, H.E.S.S., 1M2H Team,
  IKI-GW Follow-up, Fermi GBM, Pi of Sky, DWF (Deeper Wider Faster Program),
  MASTER, AstroSat Cadmium Zinc Telluride Imager Team, Swift, Pierre Auger,
  ASKAP, VINROUGE, JAGWAR, Chandra Team at McGill University, TTU-NRAO, GROWTH,
  AGILE Team, MWA, ATCA, AST3, TOROS, Pan-STARRS, NuSTAR, ATLAS Telescopes,
  BOOTES, CaltechNRAO, LIGO Scientific, High Time Resolution Universe Survey,
  Nordic Optical Telescope, Las Cumbres Observatory Group, TZAC Consortium,
  LOFAR, IPN, DLT40, Texas Tech University, HAWC, ANTARES, KU, Dark Energy
  Camera GW-EM, CALET, Euro VLBI Team, ALMA) 2017 {\em Astrophys. J.\/} {\bf
  848} L12 (\textit{Preprint} \eprint{1710.05833})

\bibitem{Arcavi:2017xiz}
Arcavi I {\em et~al.\/} 2017 {\em Nature\/} {\bf 551} 64 (\textit{Preprint}
  \eprint{1710.05843})

\bibitem{Coulter:2017wya}
Coulter D~A {\em et~al.\/} 2017 {\em Science\/} [Science358,1556(2017)]
  (\textit{Preprint} \eprint{1710.05452})

\bibitem{Lipunov:2017dwd}
Lipunov V~M {\em et~al.\/} 2017 {\em Astrophys. J.\/} {\bf 850} L1
  (\textit{Preprint} \eprint{1710.05461})

\bibitem{Soares-Santos:2017lru}
Soares-Santos M {\em et~al.\/} (DES) 2017 {\em Astrophys. J.\/} {\bf 848} L16
  (\textit{Preprint} \eprint{1710.05459})

\bibitem{Tanvir:2017pws}
Tanvir N~R {\em et~al.\/} 2017 {\em Astrophys. J.\/} {\bf 848} L27
  (\textit{Preprint} \eprint{1710.05455})

\bibitem{Valenti:2017ngx}
Valenti S, Sand D~J, Yang S, Cappellaro E, Tartaglia L, Corsi A, Jha S~W,
  Reichart D~E, Haislip J and Kouprianov V 2017 {\em Astrophys. J.\/} {\bf 848}
  L24 (\textit{Preprint} \eprint{1710.05854})

\bibitem{Abbott:2018wiz}
Abbott B~P {\em et~al.\/} (Virgo, LIGO Scientific) 2018 {\em arXiv:
  1805.11579\/} (\textit{Preprint} \eprint{1805.11579})

\bibitem{Hinderer:2018pei}
Hinderer T {\em et~al.\/} 2018  (\textit{Preprint} \eprint{1808.03836})

\bibitem{GarciaBellido:1996qt}
Garc\'{\i}a-Bellido J, Linde A and Wands D 1996 {\em Phys. Rev. D\/} {\bf
  54}(10) 6040--6058
  \urlprefix\url{https://link.aps.org/doi/10.1103/PhysRevD.54.6040}

\bibitem{Barack:2018yly}
Barack L {\em et~al.\/} 2018  (\textit{Preprint} \eprint{1806.05195})

\bibitem{Dietrich:2018jov}
Dietrich T, Day F, Clough K, Coughlin M and Niemeyer J 2018 {\em arXiv:
  1808.04746\/} (\textit{Preprint} \eprint{1808.04746})

\bibitem{Blanchet:2013haa}
Blanchet L 2014 {\em Living Rev. Relativity\/} {\bf 17} 2 (\textit{Preprint}
  \eprint{1310.1528})

\bibitem{Tagoshi:2014xsa}
Tagoshi H, Mishra C~K, Pai A and Arun K~G 2014 {\em Phys. Rev.\/} {\bf D90}
  024053 (\textit{Preprint} \eprint{1403.6915})

\bibitem{Duez:2008rb}
Duez M~D, Foucart F, Kidder L~E, Pfeiffer H~P, Scheel M~A and Teukolsky S~A
  2008 {\em Phys. Rev.\/} {\bf D78} 104015 (\textit{Preprint}
  \eprint{0809.0002})

\bibitem{Kyutoku:2010zd}
Kyutoku K, Shibata M and Taniguchi K 2010 {\em Phys. Rev.\/} {\bf D82} 044049
  [Erratum: Phys. Rev.D84,049902(2011)] (\textit{Preprint} \eprint{1008.1460})

\bibitem{Etienne:2011ea}
Etienne Z~B, Liu Y~T, Paschalidis V and Shapiro S~L 2012 {\em Phys. Rev.\/}
  {\bf D85} 064029 (\textit{Preprint} \eprint{1112.0568})

\bibitem{Kyutoku:2011vz}
Kyutoku K, Okawa H, Shibata M and Taniguchi K 2011 {\em Phys. Rev.\/} {\bf D84}
  064018 (\textit{Preprint} \eprint{1108.1189})

\bibitem{Foucart:2012vn}
Foucart F, Deaton M~B, Duez M~D, Kidder L~E, MacDonald I {\em et~al.\/} 2013
  {\em Phys.Rev.\/} {\bf D87} 084006 (\textit{Preprint} \eprint{1212.4810})

\bibitem{Foucart:2013psa}
Foucart F, Buchman L, Duez M~D, Grudich M, Kidder L~E {\em et~al.\/} 2013 {\em
  Phys.Rev.\/} {\bf D88} 064017 (\textit{Preprint} \eprint{1307.7685})

\bibitem{Foucart:2014nda}
Foucart F, Deaton M~B, Duez M~D, O'Connor E, Ott C~D, Haas R, Kidder L~E,
  Pfeiffer H~P, Scheel M~A and Szilagyi B 2014 {\em Phys. Rev.\/} {\bf D90}
  024026 (\textit{Preprint} \eprint{1405.1121})

\bibitem{Kawaguchi:2015bwa}
Kawaguchi K, Kyutoku K, Nakano H, Okawa H, Shibata M and Taniguchi K 2015 {\em
  Phys. Rev.\/} {\bf D92} 024014 (\textit{Preprint} \eprint{1506.05473})

\bibitem{Foucart:2018lhe}
Foucart F {\em et~al.\/} 2018  (\textit{Preprint} \eprint{1812.06988})

\bibitem{Bernuzzi:2014owa}
Bernuzzi S, Nagar A, Dietrich T and Damour T 2015 {\em Phys.Rev.Lett.\/} {\bf
  114} 161103 (\textit{Preprint} \eprint{1412.4553})

\bibitem{Hotokezaka:2015xka}
Hotokezaka K, Kyutoku K, Okawa H and Shibata M 2015 {\em Phys. Rev.\/} {\bf
  D91} 064060 (\textit{Preprint} \eprint{1502.03457})

\bibitem{Hinderer:2016eia}
Hinderer T {\em et~al.\/} 2016 {\em Phys. Rev. Lett.\/} {\bf 116} 181101
  (\textit{Preprint} \eprint{1602.00599})

\bibitem{Nagar:2018zoe}
Nagar A {\em et~al.\/} 2018 {\em Phys. Rev.\/} {\bf D98} 104052
  (\textit{Preprint} \eprint{1806.01772})

\bibitem{Lackey:2011vz}
Lackey B~D, Kyutoku K, Shibata M, Brady P~R and Friedman J~L 2012 {\em
  Phys.Rev.\/} {\bf D85} 044061 (\textit{Preprint} \eprint{1109.3402})

\bibitem{Lackey:2013axa}
Lackey B~D, Kyutoku K, Shibata M, Brady P~R and Friedman J~L 2014 {\em
  Phys.Rev.\/} {\bf D89} 043009 (\textit{Preprint} \eprint{1303.6298})

\bibitem{Pannarale:2013uoa}
Pannarale F, Berti E, Kyutoku K and Shibata M 2013 {\em Phys. Rev.\/} {\bf D88}
  084011 (\textit{Preprint} \eprint{1307.5111})

\bibitem{Pannarale:2015jia}
Pannarale F, Berti E, Kyutoku K, Lackey B~D and Shibata M 2015 {\em Phys.
  Rev.\/} {\bf D92} 081504 (\textit{Preprint} \eprint{1509.06209})

\bibitem{Pannarale:2015jka}
Pannarale F, Berti E, Kyutoku K, Lackey B~D and Shibata M 2015 {\em Phys.
  Rev.\/} {\bf D92} 084050 (\textit{Preprint} \eprint{1509.00512})

\bibitem{Barnes:2016umi}
Barnes J, Kasen D, Wu M~R and Martínez-Pinedo G 2016 {\em Astrophys. J.\/}
  {\bf 829} 110 (\textit{Preprint} \eprint{1605.07218})

\bibitem{Fernandez:2016sbf}
Fernández R, Foucart F, Kasen D, Lippuner J, Desai D and Roberts L~F 2017 {\em
  Class. Quant. Grav.\/} {\bf 34} 154001 (\textit{Preprint}
  \eprint{1612.04829})

\bibitem{Kasen:2017sxr}
Kasen D, Metzger B, Barnes J, Quataert E and Ramirez-Ruiz E 2017 {\em Nature,
  10.1038/nature24453\/} (\textit{Preprint} \eprint{1710.05463})

\bibitem{Kawaguchi:2018ptg}
Kawaguchi K, Shibata M and Tanaka M 2018 {\em Astrophys. J.\/} {\bf 865} L21
  (\textit{Preprint} \eprint{1806.04088})

\bibitem{Metzger:2016pju}
Metzger B~D 2017 {\em Living Rev. Rel.\/} {\bf 20} 3 (\textit{Preprint}
  \eprint{1610.09381})

\bibitem{Kawaguchi:2016ana}
Kawaguchi K, Kyutoku K, Shibata M and Tanaka M 2016 {\em Astrophys. J.\/} {\bf
  825} 52 (\textit{Preprint} \eprint{1601.07711})

\bibitem{Perego:2017wtu}
Perego A, Radice D and Bernuzzi S 2017 {\em Astrophys. J.\/} {\bf 850} L37
  (\textit{Preprint} \eprint{1711.03982})

\bibitem{Huang:2018vdq}
Huang Z~Q, Liu L~D, Wang X~Y and Dai Z~G 2018 {\em Astrophys. J.\/} {\bf 867} 6
  (\textit{Preprint} \eprint{1809.03087})

\bibitem{Shibata:2011jka}
Shibata M and Taniguchi K 2011 {\em Living Rev. Rel.\/} {\bf 14} 6

\bibitem{Foucart:2012nc}
Foucart F 2012 {\em Phys. Rev.\/} {\bf D86} 124007 (\textit{Preprint}
  \eprint{1207.6304})

\bibitem{Foucart:2018rjc}
Foucart F, Hinderer T and Nissanke S 2018 {\em Phys. Rev.\/} {\bf D98} 081501
  (\textit{Preprint} \eprint{1807.00011})

\bibitem{De:2018uhw}
De S, Finstad D, Lattimer J~M, Brown D~A, Berger E and Biwer C~M 2018 {\em
  Phys. Rev. Lett.\/} {\bf 121} 091102 (\textit{Preprint} \eprint{1804.08583})

\bibitem{Abbott:2018exr}
Abbott B~P {\em et~al.\/} (Virgo, LIGO Scientific) 2018 {\em {arXiv:
  1805.11581}\/} (\textit{Preprint} \eprint{1805.11581})

\bibitem{Radice:2017lry}
Radice D, Perego A, Zappa F and Bernuzzi S 2018 {\em Astrophys. J.\/} {\bf 852}
  L29 (\textit{Preprint} \eprint{1711.03647})

\bibitem{Radice:2018ozg}
Radice D and Dai L 2018 {\em arXiv:1810.12917\/} (\textit{Preprint}
  \eprint{1810.12917})

\bibitem{Coughlin:2018miv}
Coughlin M~W, Dietrich T, Doctor Z, Kasen D, Coughlin S, Jerkstrand A, Leloudas
  G, McBrien O, Metzger B~D, O’Shaughnessy R and Smartt S~J 2018 {\em Monthly
  Notices of the Royal Astronomical Society\/} {\bf 480} 3871--3878
  (\textit{Preprint} \eprint{1805.09371})

\bibitem{Bauswein:2017vtn}
Bauswein A, Just O, Janka H~T and Stergioulas N 2017 {\em Astrophys. J.\/} {\bf
  850} L34 (\textit{Preprint} \eprint{1710.06843})

\bibitem{Annala:2017llu}
Annala E, Gorda T, Kurkela A and Vuorinen A 2018 {\em Phys. Rev. Lett.\/} {\bf
  120} 172703 (\textit{Preprint} \eprint{1711.02644})

\bibitem{Most:2018hfd}
Most E~R, Weih L~R, Rezzolla L and Schaffner-Bielich J 2018 {\em Phys. Rev.
  Lett.\/} {\bf 120} 261103 (\textit{Preprint} \eprint{1803.00549})

\bibitem{Ruiz:2017due}
Ruiz M, Shapiro S~L and Tsokaros A 2018 {\em Phys. Rev.\/} {\bf D97} 021501
  (\textit{Preprint} \eprint{1711.00473})

\bibitem{Margalit:2017dij}
Margalit B and Metzger B~D 2017 {\em Astrophys. J.\/} {\bf 850} L19
  (\textit{Preprint} \eprint{1710.05938})

\bibitem{Rezzolla:2017aly}
Rezzolla L, Most E~R and Weih L~R 2018 {\em Astrophys. J.\/} {\bf 852} L25
  (\textit{Preprint} \eprint{1711.00314})

\bibitem{Shibata:2017xdx}
Shibata M, Fujibayashi S, Hotokezaka K, Kiuchi K, Kyutoku K, Sekiguchi Y and
  Tanaka M 2017 {\em Phys. Rev.\/} {\bf D96} 123012 (\textit{Preprint}
  \eprint{1710.07579})

\bibitem{Coughlin:2018fis}
Coughlin M~W, Dietrich T, Margalit B and Metzger B~D 2018  (\textit{Preprint}
  \eprint{1812.04803})

\bibitem{Ashton:2018jfp}
Ashton G {\em et~al.\/} 2019 {\em Astrophys. J. Suppl.\/} {\bf 241} 27
  (\textit{Preprint} \eprint{1811.02042})

\bibitem{Dietrich:2018uni}
Dietrich T {\em et~al.\/} 2018 {\em arXiv: 1804.02235\/} (\textit{Preprint}
  \eprint{1804.02235})

\bibitem{Hjorth:2017a}
Hjorth J, Levan A~J, Tanvir N~R, Lyman J~D, Wojtak R, Schr{\o}der S~L, Mandel
  I, Gall C and Bruun S~H 2017 {\em The Astrophysical Journal\/} {\bf 848} L31
  \urlprefix\url{https://doi.org/10.3847%2F2041-8213%2Faa9110}

\bibitem{vanEerten:2018vgj}
van Eerten E~T~H, Ryan G, Ricci R, Burgess J~M, Wieringa M, Piro L, Cenko S~B
  and Sakamoto T 2018 {\em arXiv:1808.06617\/} (\textit{Preprint}
  \eprint{1808.06617})

\bibitem{Strobel:1999vn}
Strobel K, Schaab C and Weigel M~K 1999 {\em Astron. Astrophys.\/} {\bf 350}
  497 (\textit{Preprint} \eprint{astro-ph/9908132})

\bibitem{Suwa:2018uni}
Suwa Y, Yoshida T, Shibata M, Umeda H and Takahashi K 2018  (\textit{Preprint}
  \eprint{1808.02328})

\bibitem{Smartt:2017fuw}
Smartt S~J {\em et~al.\/} 2017 {\em Nature, 10.1038/nature24303\/}
  (\textit{Preprint} \eprint{1710.05841})

\bibitem{Kohri:2005tq}
Kohri K, Narayan R and Piran T 2005 {\em Astrophys. J.\/} {\bf 629} 341--361
  (\textit{Preprint} \eprint{astro-ph/0502470})

\bibitem{Surman:2005kf}
Surman R, McLaughlin G~C and Hix W~R 2006 {\em Astrophys. J.\/} {\bf 643}
  1057--1064 (\textit{Preprint} \eprint{astro-ph/0509365})

\bibitem{Metzger:2008av}
Metzger B, Piro A and Quataert E 2008 {\em Mon.Not.Roy.Astron.Soc.\/} {\bf 390}
  781 (\textit{Preprint} \eprint{0805.4415})

\bibitem{Siegel:2014ita}
Siegel D~M, Ciolfi R and Rezzolla L 2014 {\em Astrophys. J.\/} {\bf 785} L6
  (\textit{Preprint} \eprint{1401.4544})

\bibitem{Just:2014fka}
Just O, Bauswein A, Pulpillo R~A, Goriely S and Janka H~T 2015 {\em Mon. Not.
  Roy. Astron. Soc.\/} {\bf 448} 541--567 (\textit{Preprint}
  \eprint{1406.2687})

\bibitem{Rezzolla:2014nva}
Rezzolla L and Kumar P 2015 {\em Astrophys. J.\/} {\bf 802} 95
  (\textit{Preprint} \eprint{1410.8560})

\bibitem{Song:2017qyc}
Song C~Y, Liu T and Li A 2018 {\em Mon. Not. Roy. Astron. Soc.\/} {\bf 477}
  2173--2182 (\textit{Preprint} \eprint{1710.00142})

\bibitem{Coughlin:2017ydf}
Coughlin M, Dietrich T, Kawaguchi K, Smartt S, Stubbs C and Ujevic M 2017 {\em
  Astrophys. J.\/} {\bf 849} 12 (\textit{Preprint} \eprint{1708.07714})

\bibitem{Etienne:2008re}
Etienne Z~B, Liu Y~T, Shapiro S~L and Baumgarte T~W 2009 {\em Phys. Rev.\/}
  {\bf D79} 044024 (\textit{Preprint} \eprint{0812.2245})

\bibitem{Foucart:2010eq}
Foucart F, Duez M~D, Kidder L~E and Teukolsky S~A 2011 {\em Phys. Rev.\/} {\bf
  D83} 024005 (\textit{Preprint} \eprint{1007.4203})

\bibitem{Foucart:2011mz}
Foucart F, Duez M~D, Kidder L~E, Scheel M~A, Szilagyi B and Teukolsky S~A 2012
  {\em Phys. Rev.\/} {\bf D85} 044015 (\textit{Preprint} \eprint{1111.1677})

\bibitem{Foucart:2019bxj}
Foucart F, Duez M~D, Kidder L~E, Nissanke S, Pfeiffer H~P and Scheel M~A 2019
  {\em Phys. Rev.\/} {\bf D99} 103025 (\textit{Preprint} \eprint{1903.09166})

\bibitem{Eichler:1989ve}
Eichler D, Livio M, Piran T and Schramm D~N 1989 {\em Nature\/} {\bf 340}
  126--128

\bibitem{Paczynski:1991aq}
Paczynski B 1991 {\em Acta Astron.\/} {\bf 41} 257--267

\bibitem{Meszaros:1992ps}
Meszaros P and Rees M~J 1992 {\em Astrophys. J.\/} {\bf 397} 570--575

\bibitem{Narayan:1992iy}
Narayan R, Paczynski B and Piran T 1992 {\em Astrophys. J.\/} {\bf 395}
  L83--L86 (\textit{Preprint} \eprint{astro-ph/9204001})

\bibitem{Wu:2018bxg}
Wu Y and MacFadyen A 2018 {\em arXiv:1809.06843\/} (\textit{Preprint}
  \eprint{1809.06843})

\bibitem{Wang:2018nye}
Wang Y~Z, Shao D~S, Jiang J~L, Tang S~P, Ren X~X, Jin Z~P, Fan Y~Z and Wei D~M
  2018 {\em arXiv:1811.02558\/} (\textit{Preprint} \eprint{1811.02558})

\bibitem{PhysRevD.99.083016}
Carson Z, Chatziioannou K, Haster C~J, Yagi K and Yunes N 2019 {\em Phys. Rev.
  D\/} {\bf 99}(8) 083016
  \urlprefix\url{https://link.aps.org/doi/10.1103/PhysRevD.99.083016}

\bibitem{Dietrich:2016fpt}
Dietrich T and Ujevic M 2017 {\em Class. Quant. Grav.\/} {\bf 34} 105014
  (\textit{Preprint} \eprint{1612.03665})

\bibitem{Radice:2018pdn}
Radice D, Perego A, Hotokezaka K, Fromm S~A, Bernuzzi S and Roberts L~F 2018
  {\em arXiv: 1809.11161\/}

\bibitem{Carson:2019rjx}
Carson Z, Chatziioannou K, Haster C~J, Yagi K and Yunes N 2019 {\em Phys.
  Rev.\/} {\bf D99} 083016 (\textit{Preprint} \eprint{1903.03909})

\bibitem{Siegel:2017nub}
Siegel D~M and Metzger B~D 2017 {\em Phys. Rev. Lett.\/} {\bf 119} 231102
  (\textit{Preprint} \eprint{1705.05473})

\bibitem{Siegel:2017jug}
Siegel D~M and Metzger B~D 2018 {\em Astrophys. J.\/} {\bf 858} 52
  (\textit{Preprint} \eprint{1711.00868})

\bibitem{Fernandez:2018kax}
Fernández R, Tchekhovskoy A, Quataert E, Foucart F and Kasen D 2019 {\em Mon.
  Not. Roy. Astron. Soc.\/} {\bf 482} 3373 (\textit{Preprint}
  \eprint{1808.00461})

\end{thebibliography}

\end{document}